\documentclass{IEEEcsmag}

\usepackage[colorlinks,urlcolor=blue,linkcolor=blue,citecolor=blue]{hyperref}

\usepackage[colorinlistoftodos]{todonotes}
\usepackage{upmath}

\jvol{XX}
\jnum{XX}
\paper{8}
\jmonth{May/June}
\jname{}
\pubyear{2021}

\begin{document}

\sptitle{Department: Head}
\editor{Editor: Name, xxxx@email}

\title{Visualization Design Sprints for Online and On-Campus Courses}

\author{Johanna Beyer}
\affil{Harvard University}

\author{Yalong Yang}
\affil{Virginia Tech}

\author{Hanspeter Pfister}
\affil{Harvard University}

\markboth{}{Paper title}

\begin{abstract}
We present how to integrate \emph{Design Sprints} and project-based learning into introductory visualization courses. A design sprint is a unique process based on rapid prototyping and user testing to define goals and validate ideas before starting costly development. The well-defined, interactive, and time-constrained design cycle makes design sprints a promising option for teaching project-based and active-learning-centered courses to increase student engagement and hands-on experience. Over the past five years, we have adjusted the design sprint methodology for teaching a range of visualization courses. We present a detailed guide on incorporating design sprints into large undergraduate and small professional development courses in both online and on-campus settings. Design sprint results, including quantitative and qualitative student feedback, show that design sprints engage students and help practice and apply visualization and design skills. We provide design sprint teaching materials, show examples of student-created work, and discuss limitations and lessons learned.
\end{abstract}

\maketitle

\begin{figure*}[tb]
 	\centering
 	\includegraphics[width=0.9\linewidth]{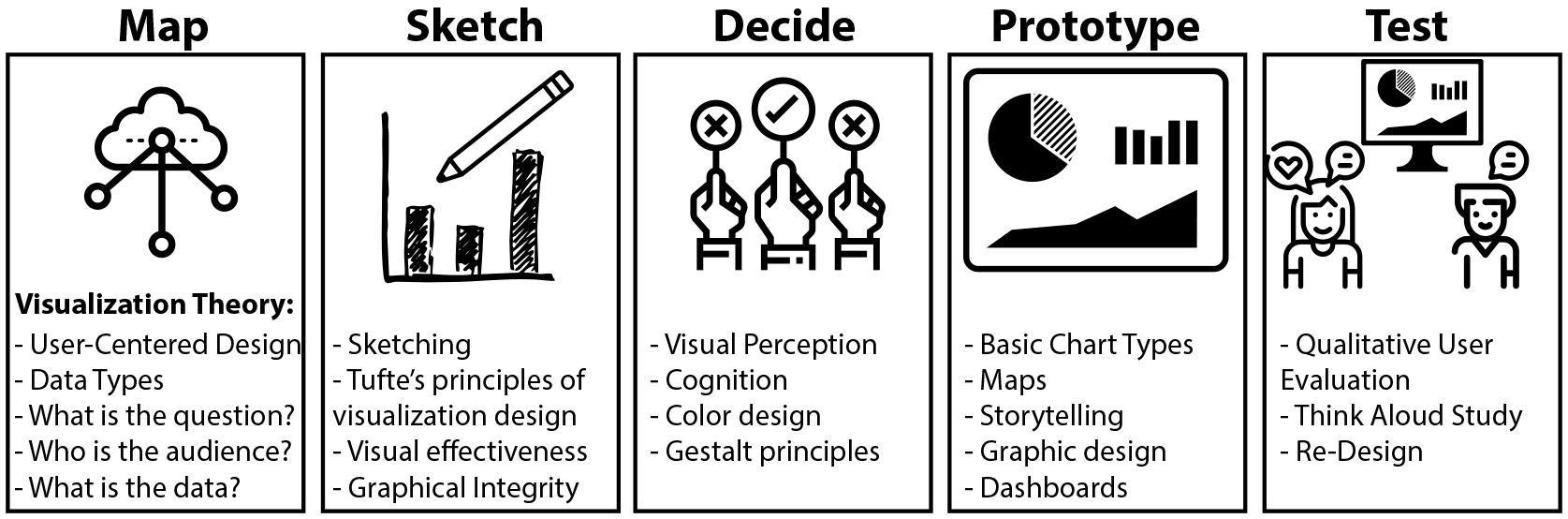}
 	\caption{The design sprint process (map, sketch, decide, prototype, test) and the main visualization-related activities in each step.}
 	\label{fig:design_sprint_steps}
 	\vspace{-4mm}
\end{figure*}

\chapterinitial{Innovative pedagogical techniques} such as active learning and flipped classrooms have become increasingly popular in recent years for teaching university-level courses.
Active learning encourages students to become actively engaged in higher-order thinking tasks, such as analysis, synthesis, and evaluation~\cite{1_princeDoesActiveLearning2004}.
Additionally, learner-centered principles and group-based projects help students to develop problem-solving skills and foster collaborative learning.

Recently, the concept of design sprints has been explored in classroom settings~\cite{2_ferreiraDesignSprintClassroom2019}, as they inherently support active learning and learner-centered teaching methodologies. Design sprints are time-constrained, interdisciplinary projects that rely heavily on rapid prototyping and testing with customers to quickly define goals and validate ideas~\cite{3_knappSprintHowSolve2016}. Design sprints consist of the five distinct steps shown in \textbf{Fig.~\ref{fig:design_sprint_steps}}.

This paper outlines how we have incorporated and adjusted design sprints for college-level and professional education data visualization courses.
This work extends our previous workshop presentation on visualization design sprints~\cite{4_beyer20} to include more details on visualization-specific adjustments of design sprints and their translation to online teaching.
We use design sprints in our classes to teach students visualization skills, a structured user-centered design process, and the necessary skills for team-based projects.
Using active learning with in-class activities and multi-week group projects leaves less time for content coverage. Instead, our students gain hands-on experience and can apply their knowledge in a real-world project, leading to a noticeable improvement in student project quality.

Starting in spring 2016, we transformed our undergraduate visualization course at Harvard University with more than 200 students from lecture-centered to learner-centered pedagogy~\cite{5_beyerTeachingVisualizationLarge2016}.
We gradually added design sprints to teach a user-centered design process and conduct team-based visualization projects~\cite{6_hilburnTeachingTeamwork2002}.

Over the past five years, we have used design sprints in our semester-long visualization course CS171 (\url{www.cs171.org}) and several two-day visualization workshops for professionals (\url{www.dataviscourse.org}). In total, we have used design sprints 11 times on-campus and ten times online for a total of 21 times in four different settings (college and professional education, both on-campus and online, respectively). In addition to adjusting the length of the design sprint, we also introduced visualization components at each step (\textbf{Fig.~\ref{fig:design_sprint_steps}, bottom}).

In the remainder of this paper, we provide a detailed outline of how we conduct design sprints in our visualization courses. We show several examples of how to incorporate design sprints into different course settings. Along the way, we discuss common pitfalls, limitations and present several lessons learned. Our teaching materials are freely available on GitHub (\url{https://github.com/CS171/design_sprint_material}), and we give guidance on how to prepare your own teaching materials. We believe that design sprints are valuable for active learning and project-based courses that prioritize hands-on visualization skills.

\section{DESIGN SPRINT OVERVIEW}
\label{sec:sprints}

The original design sprint is a five-day process that focuses on rapid prototyping and customer feedback to generate ideas and solve challenging problems. Design sprints were developed at Google Ventures and focus on rapidly exploring a business idea in one 40-hour workweek~\cite{3_knappSprintHowSolve2016}. A design sprint team should be interdisciplinary and consist of people from different fields such as engineering, design, marketing, or finance. Each team also needs a \emph{decider} with the final say in decisions (e.g., a CEO) and a \emph{facilitator} responsible for keeping track of time and organizing the sprint process.

A design sprint consists of five distinct steps: Map, Sketch, Decide, Prototype, and Test. It starts with mapping out the problem space, then sketching potential solutions, deciding on one of them, and finally prototyping and testing the chosen approach.

While a design sprint does not end in a finished product, it is an efficient way to validate product ideas.
Compared to other strategies for quick prototyping, such as hackathons, design sprints focus on real users' needs and follow a clearly defined five-step user-centered process that naturally lends itself to teaching.

\section{COURSE STRUCTURE}
\label{sec:course_structure}

We describe our learning goals and course structure before outlining how we adjusted each design sprint step in our visualization courses.

\subsection{Learning Goals}
\begin{figure}[tb]
 	\centering
 	\includegraphics[width=1.0\columnwidth]{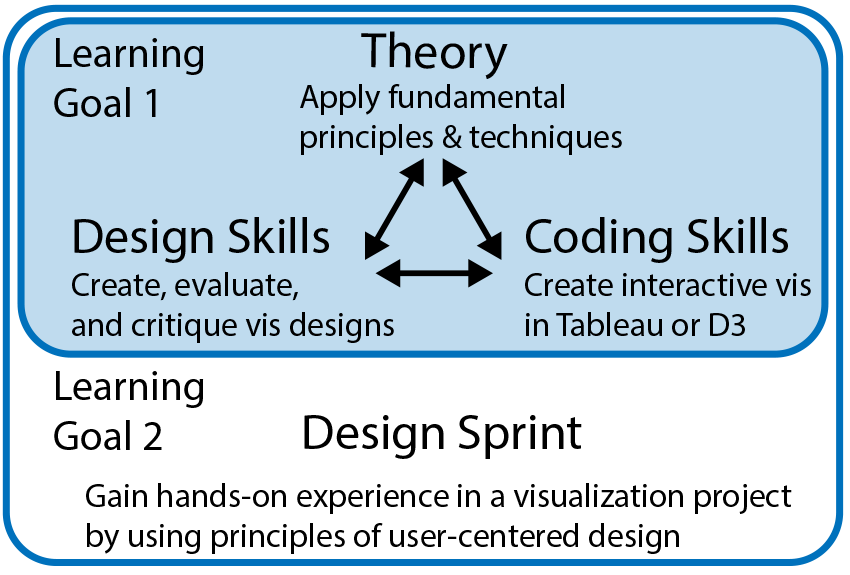}
 	\vspace{-6mm}
 	\caption{Course learning goals. Design sprints allow students to apply their knowledge of visualization theory, design, and coding skills to gain hands-on experience in visualization projects using user-centered design principles.}
 	\label{fig:goals}
 	\vspace{-4mm}
\end{figure}

Our design sprints' learning goals are twofold (\textbf{Fig.~\ref{fig:goals}}): First, we want students to practice the triad of visualization skills they have learned in class (i.e., visualization theory, design skills, and coding skills) by applying them to more substantial projects. Second, we want students to learn how to conduct team-based visualization projects with a user-centered design process.
Visualization design sprints naturally combine the three main themes of our first learning goal: First, students learn about visualization theory and how to apply those fundamental principles and techniques to their project.
Since we teach introductory visualization courses, we focus on high-level concepts of visualization theory, including Tufte's design principles, Gestalt principles, basics of human visual perception and cognition, basic chart types, color, and storytelling.
Second, students learn design skills and how to create, evaluate, and critique visualizations. We cover sketching, peer reviews, and in-class discussion of visualization case studies. Third, students learn the technical skills needed to create interactive data visualizations. Depending on the length of the course, this includes using tools like Tableau or programming in JavaScript and D3.
While we cover each of these elements in other parts of our teaching (e.g., programming labs, lectures, design exercises), design sprints allow students to iterate on the covered concepts and to learn by doing.

To achieve our second learning goal, we introduce students to the user-centered design process and teach proper teamwork and management. Students learn to start with the user and the user's needs and develop their ideas based on their user in mind.

\begin{figure*}[tb]
 	\centering
 	\includegraphics[width=0.99\linewidth]{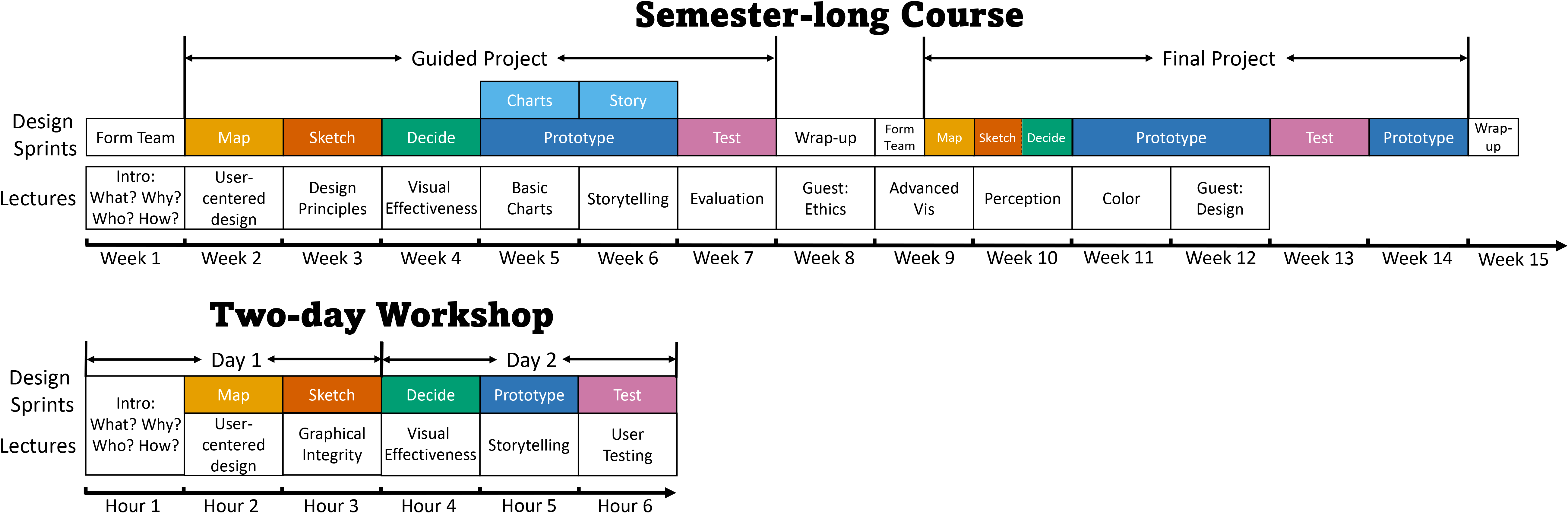}
 	\caption{Our schedule for design sprints in semester-long courses and two-day workshops. Design sprint steps are highlighted in color, and lecturing is interspersed with design sprint work.}
 	\label{fig:schedule}
 	\vspace{-4mm}
\end{figure*}

\subsection{Design Sprint Types}
In our courses, we run two different types of design sprints. The first type is a \emph{guided project} that acts as a scaffolded introduction to the design sprint process. In our college visualization course, we run the guided project in the first half of the semester and use the last 20 minutes of each lecture for design sprint activities. Each step of the guided sprint takes approximately one week.
In our professional education workshops, we run a guided sprint throughout the entire workshop and intersperse it with more traditional lectures and shorter activities, as shown in \textbf{Fig.~\ref{fig:schedule}}.
This integration of the design sprint into class time allows us to give immediate feedback while teaching students the design sprint process.

The second type of design sprint we have incorporated into our college course is a \emph{final project}. Students work on the final project mostly outside of class and hand in the result at the end of the semester. Each step in the final project still takes approximately one week, except for the prototyping phase. Since we expect a working interactive D3 visualization at the end of the final project, we allot three weeks for prototyping and necessary design iterations. Overall, the final projects run for approximately six weeks.
\textbf{Fig.~\ref{fig:schedule}} shows the detailed schedule of the design sprint steps in our courses.

\subsection{Adjusting Scope and Level of Difficulty}
We adjust the scope and difficulty of our design sprints based on the students (e.g., undergraduate, graduate, or professional), their technical background (e.g., computer science or non-technical), and the length of the course (e.g., semester or two days).

One way to vary the difficulty of a design sprint is the amount of guidance that we provide. In guided projects, we give much more immediate feedback and help than during final projects. Additionally, we can adjust the level of difficulty by changing the tools used for prototyping. For example, in our college course, we require students to implement their final projects in D3, whereas we allow them to use Tableau for the guided project. Tableau is less flexible and allows less creativity. Still, students can immediately apply their knowledge on how to design visualizations without having to learn JavaScript and the intricacies of D3.
For less technical students, we can allow them to prototype their projects entirely with pen and paper sketches. Finally, we can further simplify a design sprint by giving students a pre-defined topic, audience, and clean data, as opposed to letting them choose their own.

\subsection{Teaching Staff}
For guided projects, we found it essential that students get frequent and immediate feedback. By integrating the guided projects directly into class time, course staff can check in with each group, clarify any questions, and give feedback on design decisions and the specific design sprint steps.
Additionally, we assign \emph{project mentors} to each team. These are members of the teaching staff that are responsible for overseeing 4 to 5 project teams.
During the final project, teams have in-person check-ins with their project mentor every 2-3 weeks and get weekly written feedback on their progress.
We found that having a project mentor is essential for providing consistent feedback throughout a design sprint that lasts several weeks. In larger online classes or in settings with less teaching staff, we also ask students to give frequent feedback to each other.

\begin{figure*}[tb]
 	\centering
 	\includegraphics[width=1.0\linewidth]{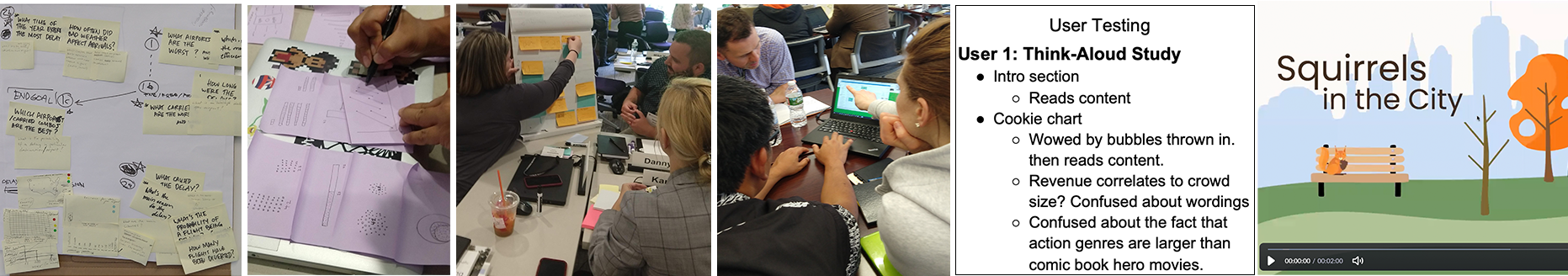}
 	\caption{Examples from visualization design sprints (from left to right): mapping out the problem space by arranging sticky notes, individual sketching, deciding on an approach by group-based voting, prototyping, user testing with a think-aloud study, and presenting a final project video.}
 	\label{fig:vis_design_sprint}
 	\vspace{-4mm}
\end{figure*}

\subsection{Assessments}
In all our design sprints, we track students' progress through weekly submissions. Weekly submissions ensure that project teams stay on track and do not fall behind, and they allow the teaching staff to provide timely written feedback.
Additionally, we include peer feedback into the design sprint process. This includes feedback from one team to another, for example, during the user testing phase. It also includes self and peer evaluations of students within each team.
If there are strong discrepancies between a students' self-evaluation and their team members' ratings, we reach out to the student and discuss possible causes and remedies. During the guided projects, these evaluations are formative and not graded. In the final projects, they can affect a students' individual project grade. Since introducing self and peer evaluations for teams, we had to deal with much fewer team performance issues.

Finally, as an additional encouragement, we award best project prizes for the top three final projects at the end of the semester. These prizes do not influence students' grades. Nevertheless, students love having bragging rights and being added to the hall of fame on our course website. They also serve as a motivating factor for project mentors, who naturally want the best project team to be one of their mentored teams.

\subsection{Online Teaching Considerations}
We have taught several design sprints successfully online, both in our semester-long courses and two-day workshops.
The main pitfall for online design sprints is that students need regular and immediate feedback. It is naturally more difficult for them to ask questions after class or drop by the teaching staff's office. Therefore, we schedule all online design sprint sessions to be synchronous during class time while teaching staff gives feedback and answers questions.

Furthermore, we slightly reduce the content for online courses, considering that teamwork and group activities take longer in online settings. Especially the two-day online workshops require careful planning and scheduling. We give students a clean dataset and also allow them to prototype solely with pen and paper sketches instead of Tableau or D3. This approach enables students without technical expertise to be highly productive in a limited amount of time.

Finally, we have found it very important for online courses to have clear, written instructions available to students prior to class. These instructions introduce the collaborative tools that will be used during the design sprint. We also highlight deadlines and outline the concrete steps and requirements for each part of the design sprint.
In our online courses, we typically use Zoom and Zoom breakout rooms for video conferencing, Google Docs to document each step of the sprint, and Google Jamboard for collaborative sketching. We limit the number of different tools that we introduce to avoid overloading students with unnecessary complexity.

\section{Visualization Design Sprints}
\label{sec:vis_sprints}
We follow the main five steps of a traditional design sprint in our courses but have added several modifications specific to visualization design sprints in classroom settings. \textbf{Fig.~\ref{fig:vis_design_sprint}} shows real-world examples of what each step in a typical visualization design sprint looks like.
One main adjustment is that we emphasize the importance of iterations in visualization design. Therefore, we typically give students time to refine their initial ideas after having received feedback from peers and course staff.
We also emphasize visual storytelling and ask students to combine individual visualizations to tell an engaging story.
Furthermore, based on our learning goals, we cover visualization theory, design skills, and visualization prototyping tools such as Tableau and D3. Here we outline how we have adjusted traditional design sprints to fit classroom settings and how we incorporate essential visualization elements.

\subsection{Design Sprint Preparation}
Before starting a design sprint in any of our courses, we first outline the format and, more importantly, explain the motivation of following the design sprint process. This introduction primes students for their projects.
Next, we start with our team formation process.
We randomly assign students to teams for all of our guided projects to create diverse groups of students with different design and coding backgrounds. Additionally, in college courses, we make sure to talk about effective teamwork. We ask students to sign \emph{team expectation agreements} that outline their responsibilities, such as their method of communication (e.g., Slack or e-mail), team meeting times, and tools for sharing materials and code. We found that this initial team formation step is crucial for setting up a positive team dynamic that prevents many issues student teams often face.
Finally, every team creates a \emph{process book}, which is a shared digital lab notebook (we use Google Docs) to document every step of the design sprint in great detail. Process books are part of the design sprint deliverables and allow course staff to give direct feedback to all teams on an ongoing basis, which is especially crucial in online courses.
It is also essential that team members divide their responsibilities before starting the design sprint. This includes selecting a team leader (i.e., a decider) and a person responsible for submitting the material to the course staff each week.

\subsection{Map}
The first step in a design sprint focuses on mapping out the problem space. This includes setting a long-term goal, coming up with questions that need to be answered to achieve this goal, and turning the goal into actionable items. It also involves getting input from experts and doing background research.

In our courses, we introduce the user-centered design process~\cite{7_lowdermilkUserCenteredDesign2013} and different data types (nominal, ordinal, and quantitative) based on Steven's taxonomy~\cite{8_stevensTheoryScalesMeasurement1946}. For guided projects, we let students select one of several clean datasets that they will be using during the sprint.
We found that preparing so-called \emph{project briefs} helps students understand the context, audience, and specific data types for each dataset. For final projects, we let students pick their own topic and data.
In their teams, they then focus on three main questions: \emph{What is the data?}, \emph{What is the question?}, and \emph{Who is the audience?}.

Students perform initial data exploration, typically using Tableau, and then discuss the potential audience and interesting questions in their groups, either in person or via video conferencing. In-person, we provide students with pens, sticky notes, and a whiteboard to jot down and arrange their ideas quickly.
Students document their progress in their online process book.

\subsection{Sketch}
This step exposes students to the ideation and divergence process by sketching many possible visualizations. We introduce students to sketching with pen and paper, illustrate Tufte's design principles~\cite{9_tufteVisualDisplayQuantitative2001}, and discuss visual effectiveness~\cite{10_mackinlayAutomatingDesignGraphical1986}.

As opposed to the other steps in the design sprint, sketching is done \emph{individually} to quickly create many diverging ideas that answer questions about the data that students have identified in the map step.
After sketching many potential solutions, each person can pick their favorite sketch and refine it to add more detail.
We ask students to use sticky notes and Sharpie pens for sketching. For in-class settings, this allows students to arrange their sketches on a whiteboard later. For online courses, Sharpies are easily visible even when sharing sketches via webcam. Alternatively, if online students have a tablet with a digital pencil, we encourage them to sketch digitally with a drawing app of their choice or Google Jamboard. At the end of this step, students take pictures or screenshots of their sketches and upload them into their process book.

\subsection{Decide}
In the third step, the team works on reaching a consensus on the best solution.
Students review all proposed sketches of their group, critique them, and vote to pick the most promising designs. We discuss human visual perception and cognition, principles of color design, the Gestalt principles~\cite{11_koffkaPrinciplesGestaltPsychology1935}, and creative considerations that might lead to breaking design guidelines.

Students start by collecting all of their sketches into a shared space (e.g., a whiteboard or Google Jamboard). They then cluster their sketches based on what questions the visualizations answer. To identify the top three to four sketches, each student votes for their favorites by placing up to three dot stickers (or their electronic initials) next to their favorites. They are allowed to place multiple votes on the same sketch and to vote for their own sketches.

After the vote, students combine the highest-rated sketches into an initial sketch of a dashboard. We explicitly encourage students to iterate over their sketches and to improve them based on the visualization principles they have learned so far. To sketch the dashboard, students use sticky notes and a whiteboard or Google Jamboard.

\subsection{Prototype}
In this step, the team builds a prototype of their idea. In the original design sprint method, this step takes one day and uses paper mock-ups and prototyping design tools.

In our courses, we use Tableau, D3, or sketching and give students more time to finish prototyping in those tools. We discuss basic chart types, maps, and visualizations for high-dimensional data. We also teach them basic graphic design principles~\cite{12_williamsNondesignerDesignBook1994} and how to tell effective data stories with visualizations.
Prior to the prototyping step, we make sure that students are sufficiently competent using prototyping tools. In our on-campus courses, we introduce Tableau and teach students to implement basic chart types and maps. In CS171, we use self-guided weekly programming labs to teach students the basics of web programming with HTML, CSS, Javascript, and D3. Since we have students with varying levels of programming experience, we dedicate one class period each week to the labs. We also add weekly programming assignments that students complete individually. Even though it takes time and effort to teach D3, we found that even students with non-technical backgrounds really like learning this skill.

During the prototyping phase, students work on their Tableau dashboards or interactive websites and data stories.
In all our courses, the prototyping step takes the longest of the design sprint steps. In semester-long courses, we typically allocate 2-3 weeks for prototyping, depending on whether it is the guided or the final project.

\subsection{Test}
The final step in a design sprint focuses on user testing. The goal is to test the prototype with 2-5 users and to collect their feedback in user interviews.
During this step, we present qualitative testing and evaluation methods.
We do not have time to cover quantitative methods.

Each group performs a qualitative user evaluation through a think-aloud study with students from other teams. First, each group decides on one team member who will test another group's project. Then, all groups run their think-aloud studies in parallel. For on-campus classes, testers simply switch tables. In online courses, they switch to a different team's breakout room. Additionally, in online courses, teams have to ensure that testers have access to the visualization project they are testing, either by hosting the project online or by sending them their Tableau files.
Students then analyze the feedback and come up with possible improvements to their initial prototypes. For final projects, teams then spend another week implementing these improvements.

\subsection{Wrap-up}
Unlike the typical design sprint process, we have added a wrap-up step to our design sprint process where teams present their projects to the class. In our semester-long course, we also ask students to fill out self and peer evaluations to rate their own and their teammates' performance during the design sprint.
For final projects, this step also involves cleaning up the process book and producing a 2-minute video of their project.
During final project presentations, we hand out best final project prizes for the top three teams as determined by the teaching staff.

\section{CASE STUDIES}

We now discuss case studies of design sprints performed in a variety of different visualization courses.
The steps of our visualization design sprints stay the same for all the different types of courses we teach (e.g., on-campus vs. online, large undergraduate courses vs. small professional development workshops).
However, we may make small adjustments in each course to fit the course objectives and available time.

We have collected datasets, teaching materials, and grading guidelines that we have curated over the past couple of years on our website (\url{https://github.com/CS171/design_sprint_material}). We also include examples of student-created process books.

\begin{figure}[tb]
    \centering
    \includegraphics[width=1.0\linewidth]{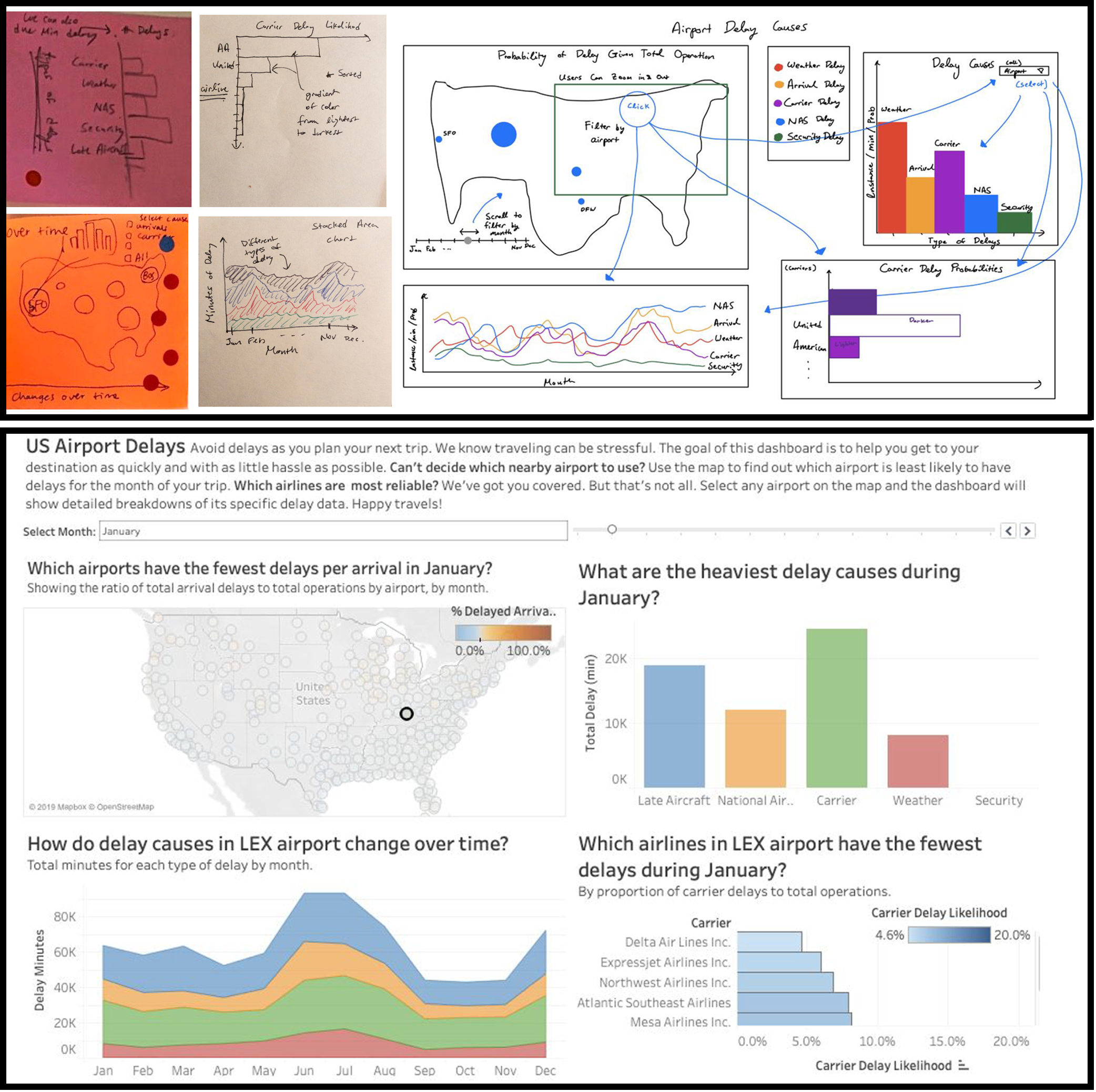}
    \vspace{-6mm}
    \caption{CS171 guided project in Tableau using U.S. flight delays data. Top left: Initial sketches; Top right: Dashboard sketch; Bottom: Tableau prototype.}
    \vspace{-4mm}
    \label{fig:case_study_guided}
\end{figure}

\noindent
\subsection{Guided Projects in Tableau}
In our semester-long courses, we start our guided sprint in the second week of the semester. We assign students randomly into teams, aiming for diverse teams in terms of programming and design experience. For in-person teaching, we use a classroom with a flat floor and movable tables so that all team members can sit together. For online teaching, we use one virtual breakout room per team.
We prepare project briefs for one to four different datasets that each team can choose from. The project briefs include a dataset description, define the audience for the final visualization prototype, and state high-level goals.
Each week, students work on one step of the design sprint and document their progress in their process book. Students get clear instructions at the beginning of each design sprint session, which we usually run in the last 20 minutes of class time, and should complete most of the work during class time. If necessary, they finish their work at home. In class, several teaching assistants walk around (or move between breakout rooms) and give feedback to groups. In addition, we hand out detailed written instructions that describe what students have to submit by the end of the week.
\textbf{Fig.~\ref{fig:case_study_guided}} shows a design sprint of flight delay data. The team created a dashboard to help casual travelers minimize flight delays. The different views are linked to each other and follow an overview-and-detail navigation pattern.

\subsection{Final Projects in D3}
\begin{figure}[tb]
    \centering
    \includegraphics[width=1.0\linewidth]{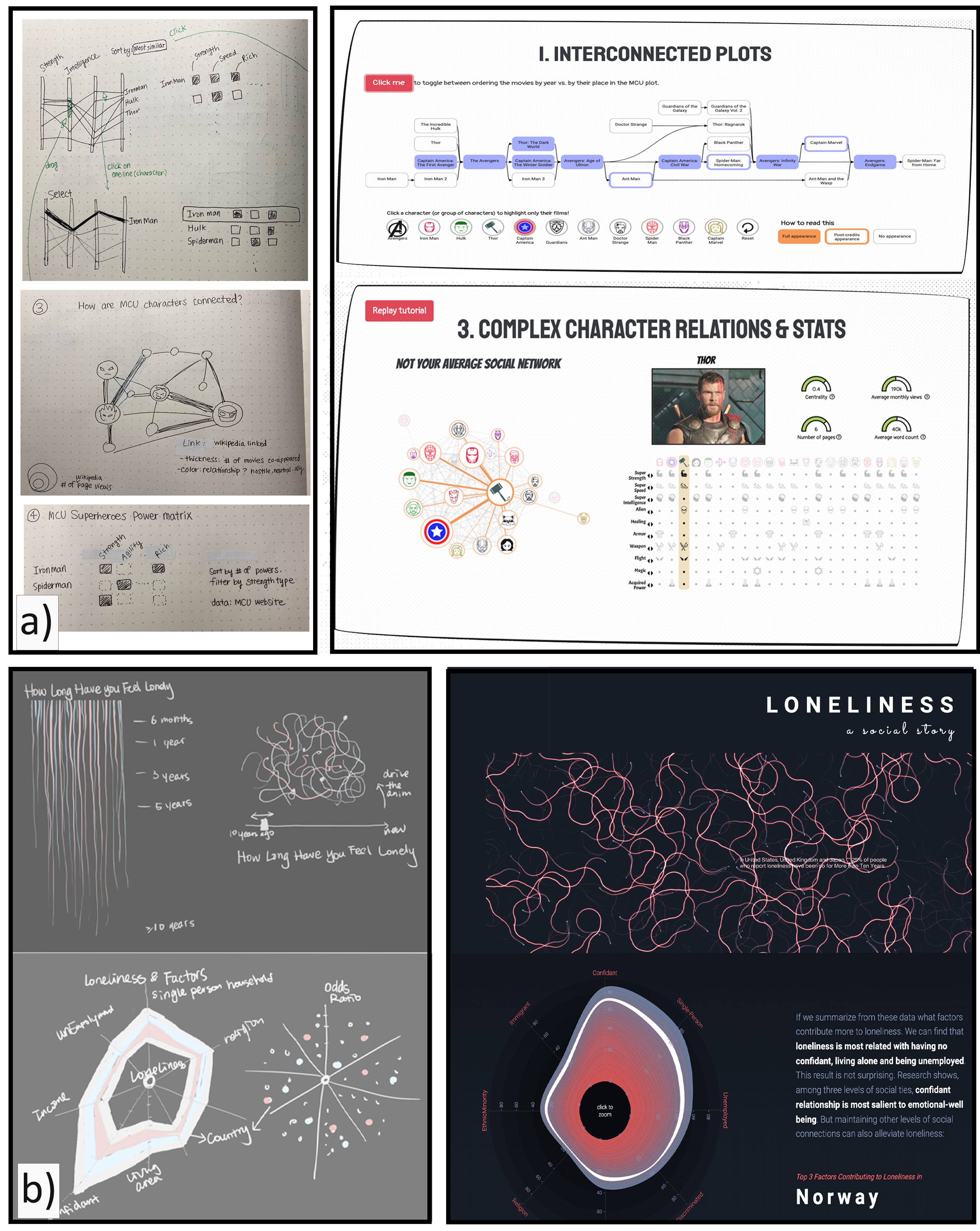}
    \vspace{-6mm}
    \caption{CS171 final projects in D3. The left column shows student sketches; the right column shows their final websites. a) Marvel superheroes. b) Project about 'Loneliness.'}
    \vspace{-4mm}
    \label{fig:case_study_final}
\end{figure}

The second design sprint in our semester-long courses is less guided and gives students more independence. Students get to pick their teams and have to find a topic and dataset they want to visualize.
Students work on their final project mainly outside of class time. However, we schedule user testing to be done during class time to circumvent scheduling issues with finding available testers.
Every week, students submit their sprint's current state and get timely feedback from their project mentor, a teaching assistant assigned to be their main point of contact for questions or feedback during the sprint.
We have observed that allowing students to pick their teams and projects leads to highly motivated student groups. This has led to a wide variety of topics over the years, including projects on obesity, the water quality of local rivers, cryptocurrency, alcohol consumption, climate change, or squirrels in the city. \textbf{Fig~\ref{fig:case_study_final}a} shows a student project on the interconnection of Marvel superheroes, and \textbf{Fig~\ref{fig:case_study_final}b} shows a project on the topic of loneliness. Both projects were done in D3 and incorporate visual storytelling and multiple connected visualizations.

Because final projects require more implementation effort than guided projects, we hold several office hours during that time to help students with technical issues in D3 and web programming.
We have also found it important to include early mechanisms of peer feedback, where students evaluate their team members. This allows us to intervene early, when necessary, and make sure that students do not fall behind.

At the end of the final project, we have a big project award session, where each group presents their project. In-person, we have a showcase where students can walk around and get demos of other students' projects. Online, we show a highlight reel of the two-minute project videos. Finally, we announce the best project teams and hand out awards, consisting of a spot on our course website's hall of fame (\url{www.cs171.org}) and Swiss chocolate.

\subsection{Two-Day Workshops}
In our two-day professional development workshops, we run an abbreviated version of our guided visualization design sprint.
Students still go through every step of the sprint but in a condensed and simplified way.
One of the main considerations for sprints in such short time spans is that we have to significantly reduce the amount of material we cover in lectures. We still introduce students to visualization concepts such as design guidelines, chart types, perception, but we do not go into as much detail.
Similarly, we shorten each design sprint step as well. We introduce students to the concepts of sprints and outline every individual step in the process, but we point out that we are doing an abbreviated form that they should extend in their future projects.
Additionally, in online workshops, we leave out the testing step due to time constraints but refer students to extra reading and resources to emphasize the importance of user testing.

\begin{figure}[tb]
    \centering
    \includegraphics[width=1.0\linewidth]{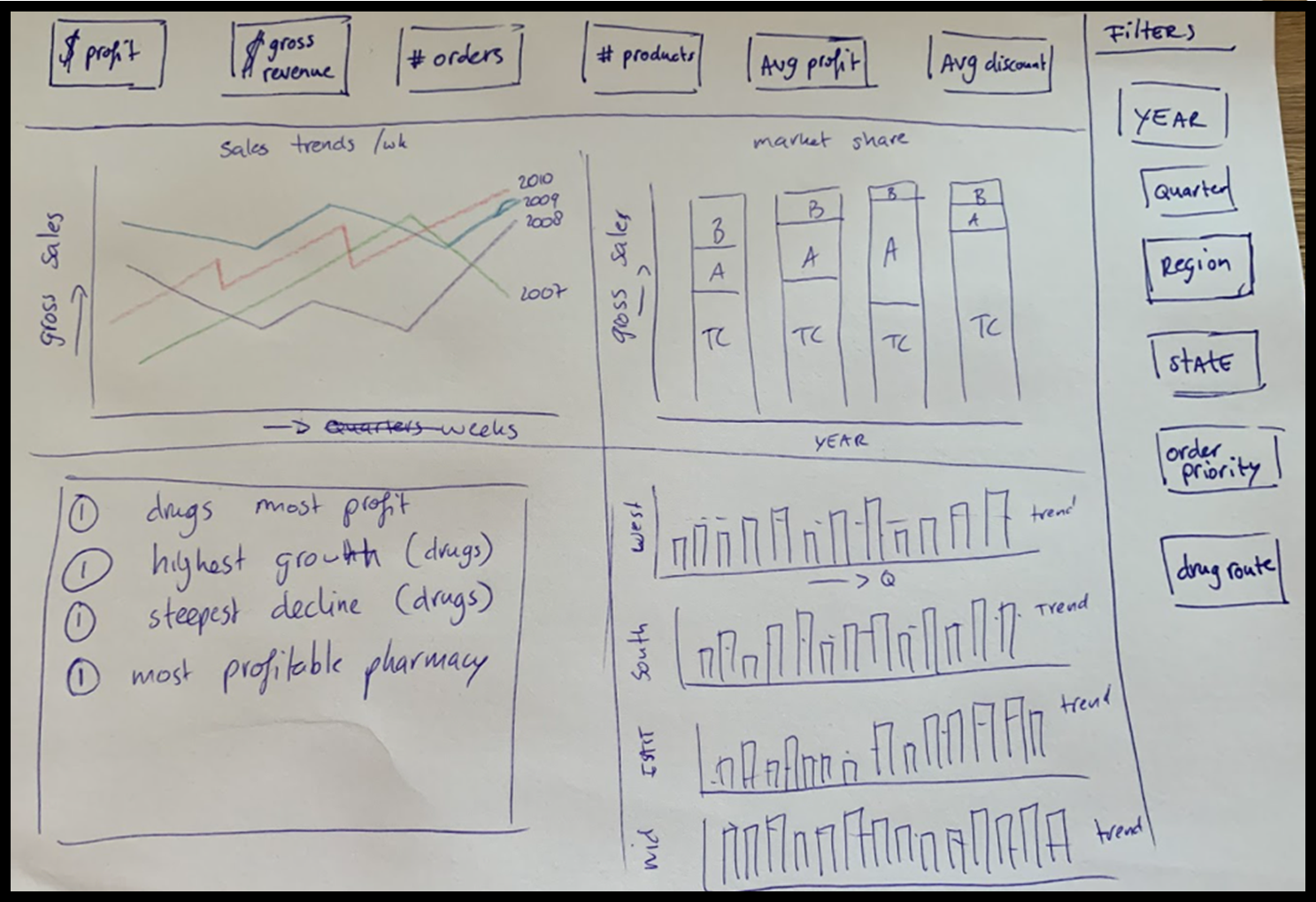}
    \vspace{-6mm}
    \caption{This dashboard design is a pen and paper sketch from a two-day online workshop, not requiring students to know Tableau or D3.}
    \vspace{-4mm}
    \label{fig:case_study_workshop}
\end{figure}

Furthermore, instead of walking around and giving immediate feedback, we found it more useful to leave written feedback in the student's online process books or Google Jamboard sketches overnight between workshop days. That gives us more time to give meaningful feedback to everyone during the in-person time. While we assign homework for each workshop day (e.g., to get acquainted with Tableau or finish their sketches), we found that only about one-third of the participants complete the assigned homework in workshop settings. In contrast to our semester-long courses, we plan the workshop such that completing the homework is not necessary for students to follow the rest of the class. This, for example, implies that some students will not have the necessary Tableau skills to create a compelling dashboard. Therefore, we give students in workshops the option to complete their prototype as pen and paper sketches (\textbf{Fig.~\ref{fig:case_study_workshop}}).

\section{DISCUSSION}
Visualization design sprints are a powerful technique to allow students to gain hands-on experience in visualization projects. We now highlight our experience with design sprints, report on student feedback, and discuss common pitfalls and limitations.

\subsection{Lessons Learned}
Design sprints allow students to apply what they have learned in class immediately. This works very well in visualization courses where students can directly apply high-level concepts, such as Tufte's design principles, to refine their visualization designs. Additionally, by using sketching and tools such as Tableau, students do not need to learn more complex visualization libraries (e.g., D3) before practicing their visualization skills. In contrast to typical design sprints, we stress the importance of iterative refinement in our visualization classes and extend the time they have to implement their prototypes.

\noindent
\textbf{Project Mentors.}
For design sprints to be effective teaching tools, students must get immediate and direct feedback. We have tackled this problem by assigning a \emph{project mentor} to each team that guides them through the design sprint and makes students aware of the strengths and weaknesses in their designs.
For example, in the two-day workshops, we read through each team's project book overnight and leave comments that students can address the next day. Being able to track student progress in a shared process book is essential, especially in the class's online formats.
Students also seem to like having a dedicated project mentor: ``\emph{Having a project mentor is a great idea –– my group learned a ton from our mentor!}''.

\noindent
\textbf{Student Engagement is High.}
We found that the optimal team size for design sprints in visualization courses is around 3-4 students. This number is big enough to ensure that students can bounce ideas off each other but small enough to keep everyone engaged and to simplify coordination.
Additionally, students will call each other out if someone is not pulling their weight.
We have tried group sizes of one to seven people, but with less consistent and positive results.
We also found that our peer feedback mechanism further helps in keeping student engagement high.

\noindent
\textbf{Project Quality Overall is High.}
In the old lecture version of the course, we noticed that projects would go wrong in the expected ways: teams missed deadlines, did not follow a structured process, ignored lessons of good visualization design, and some students were hitchhikers or couch potatoes~\cite{13_oakleyItTakesTwo2002}.
With design sprints, almost all of these problems went away. The clear and simple process makes it easier for students to know what is expected from them each week, and there is less risk of students falling behind.
We also found it helpful to share good examples from previous years with students, such as process books and prototypes.

Finally, adding a small element of competition to the design sprint (e.g., a `best project award') is appreciated by our students and our teaching fellows.
Students want to beef up their resumes, and teaching fellows also take pride in mentoring the winning team, so they are more involved during the design sprint.

\noindent
\textbf{Lessons for Online Teaching.}
An essential difference in online teaching is that students feel less connected. Therefore, clear communication is even more critical than in on-campus settings. That means that everything needs to be clearly explained and documented plenty of time ahead. For example, we clearly communicate how teams will be assembled, when and how teams are expected to meet, and we explain each step and milestone in the design sprint process. This allows online students to plan ahead and reduces student anxiety.
Networking does not work as well in online settings. Therefore, proper asynchronous communication channels (e.g., Slack or Piazza) are essential to connect students and teaching staff.
Furthermore, a video conferencing system that allows fast and easy assignment and re-assignment of breakout rooms is essential.

To avoid overwhelming students, we limit the number of different collaborative tools we use and provide usage instructions for each tool. We mainly use Google Docs and Google Jamboard for collaboration and only introduce additional tools if they add a significant benefit that cannot be achieved by video conferencing and Google Docs alone.

\noindent
\textbf{Short Online Courses.}
An important consideration for short (i.e., one- to three-day) online courses is the necessity for strict time management.
Team activities and giving feedback online take longer than in person. We found it challenging to fit a Tableau-based design sprint into our two-day online workshop. A better alternative might be to just use pen and paper prototyping in those scenarios.
Furthermore, we had to shorten and simplify lecturing content by focusing on high-level concepts such as visual effectiveness, which allowed us to combine concepts such as perception, color, and Gestalt principles into a single lecture. Furthermore, while we always teach students about user-centered design, we sometimes cut out user testing for time reasons.

The busy schedule of online workshop participants makes it hard for them to complete homework before and during the course. This makes timely and continuous feedback throughout the course even more vital.
We also found it crucial to have technical support staff to manage our online video conferencing tool, set up breakout rooms, and help students with technical issues.

\noindent
\textbf{Student feedback.}
Students have overwhelmingly liked our design sprints. For many students, the final project is the favorite part of the course. We have received feedback such as ``\emph{The final group project was a great way to combine all the skills you learned throughout the semester and turn it into some complex visualizations about real issues and topics, which is something great to be added to a portfolio.}'' and ``\emph{The final project especially showed me that I'm capable of creating unique visualizations from just the simple building blocks of D3 shapes and areas}''.

When we originally switched our large undergraduate on-campus class of roughly 200 students to active learning and a final project in 2016, our student evaluation (Q scores) increased by 0.8 points to 3.84 out of 5 (SD$=$1.12) compared to the earlier lecture-based course, which had an average Q Score of 3.0.
Since the switch to design sprints in 2018, the Q Score of CS171 has increased another 0.3 points to 4.14 (SD$=$0.83) in 2018 and 4.11 (SD$=$0.88) in 2019.
In 2020, CS171 was taught completely online due to the Covid-19 pandemic and received a Q score of 3.9 (SD$=$1.03). While this is lower than the previous two years, we see this as a natural result of online teaching and learning.
More specifically, on a 5-point Likert scale, 74\% of all students agreed or strongly agreed that the guided design sprint helped them learn how to apply visualization principles and a structured design process. 77\% of all students agreed or strongly agreed that using Tableau in the guided sprint allowed them to focus on good visualization design and visual effectiveness early on.

\subsection{Limitations}
Design sprints for teaching visualization courses also come with certain limitations.

\noindent
\textbf{Fast Results vs. Creativity.}
Originally, design sprints were proposed for prototyping in companies to quickly test product ideas within a 40-hour workweek. This focus on speed might not be ideal for teaching, where we want students to be creative, carefully consider design alternatives, and iterate on their visual designs.
To counter this, we encourage students to think outside the box and give them extra time to iterate over their designs.

Additionally, students should have the necessary programming skills to fully express themselves in their chosen visualization tools.

\noindent
\textbf{Time Constraints.}
Time constraints often require us to limit the scope and complexity of a design sprint. This comes at the risk of over-simplifying a design sprint or its data. It can be challenging to find the right balance of creating a complex but achievable design sprint. Ultimately, it might just not be feasible to run a complete design sprint in a single afternoon.

\noindent
\textbf{Scalability.}
In principle, design sprints easily scale to large classes. In practice, they do require a certain number of teaching staff.
Design sprints require frequent and timely feedback to make students aware of bad design choices and inefficient visual encodings. Peer feedback is useful in some situations (e.g., for final user testing) but less appropriate for pointing out ineffective encodings. Furthermore, giving constructive peer feedback is a skill students first need to learn.
Therefore, we recommend having at least one teaching staff per five to eight student groups.

\subsection{Guidelines in a Nutshell}
Here, we present a short list of guiding principles when running a visualization design sprint.
\begin{itemize}
    \item Introduce the design sprint process and give detailed instructions for each step.
    \item Set guidelines on how to collaborate as a team.
    \item Set clear expectations (e.g., define a rubric, provide examples of important documents and milestones).
    \item Give frequent and detailed feedback.
    \item Encourage iteration and experimentation.
    \item Include peer feedback.
    \item Allow students to pick their own project topics and keep the group size small (3-4 students).
    \item Pick visualization tools appropriate for the complexity and length of the design sprint.
    \item Do not expect students to immediately appreciate active learning and having to participate in class.
    \item Schedule a `show and tell' to allow everyone to see each other's work.
\end{itemize}

\section{CONCLUSIONS}

We will continue to use design sprints in our visualization courses, both online and in person. We believe that design sprints are an excellent technique for students to gain hands-on experience in visualization design. We distinguish between guided design sprints that are performed mainly during class time and more independent sprints that can serve as a final project.

We have noticed that design sprints lead to projects that adhere more closely to good visualization design principles, but sometimes at the cost of less creative and more uniform-looking projects. We believe that this is not necessarily due to design sprints but instead from students internalizing and applying effective visualization design principles. We counter this by explicitly encouraging students to develop novel visual designs and teach them how to implement more creative ideas in their chosen visualization software or library. We also started to emphasize more creative approaches to visualization, such as the ``Dear Data'' project~\cite{14_lupiDearData2016} or the work of Nadieh Bremer and Shirley Wu~\cite{15_datasketches} in our courses.

Finally, visualization design sprints are not just more engaging for students; they are also fun for instructors and teaching staff. Visualization design sprints have made our teaching more rewarding, and we believe that they will also benefit visualization courses at other institutions.

\bibliographystyle{unsrt}
\bibliography{designSprintActivities}

\begin{IEEEbiography}{Johanna Beyer}
is a research associate at the Visual Computing Group at Harvard University. Before joining Harvard, she was a postdoctoral fellow at the Visual Computing Center at KAUST. She received her Ph.D. in computer science at the University of Technology Vienna, Austria, in 2009. Her research interests include scalable methods for visual abstractions, large-scale volume visualization, and immersive analytics. Contact her at jbeyer@g.harvard.edu.
\end{IEEEbiography}

\begin{IEEEbiography}{Yalong Yang}
is an Assistant Professor in the Department of Computer Science at Virginia Tech. He was a Postdoctoral Fellow at the Visual Computing Group at Harvard University, and a Ph.D. student at Monash University, Australia. In his research, he designs and evaluates interactive visualizations on both conventional 2D screens and in 3D immersive environments (VR/AR).
\end{IEEEbiography}

\begin{IEEEbiography}{Hanspeter Pfister}
is An Wang Professor of Computer Science in the John A. Paulson School of Engineering and Applied Sciences at Harvard University. He has a Ph.D. in Computer Science from the State University of New York at Stony Brook and an M.S. in Electrical Engineering from ETH Zurich, Switzerland.
His research in visual computing lies at the intersection of scientific visualization, information visualization, computer graphics, and computer vision.
\end{IEEEbiography}

\end{document}